\documentstyle[12pt,amstex,epsfig]{article}
\begin{document}

\begin{flushright}
MPI-PhT/98-71\\
July 1998\\
\end{flushright}
\vskip3cm
\begin{center}
\large {\bf Possible Nonstandard Effects in $Z\gamma$ Events at LEP2}\\
\mbox{ }\\
\normalsize
\vskip1cm
{\bf K.J. Abraham}
\vskip0.3cm
Dept. of Physics \& Astronomy, Iowa State  University \\
Ames IA 50011 USA\\
\vskip1cm
{\bf Bodo Lampe}               
\vskip0.3cm
Max Planck Institut f\"ur Physik \\
F\"ohringer Ring 6, D-80805 M\"unchen \\
\vspace{1cm}

{\bf Abstract}\\
\end{center}
We point out that the so--called 'radiative return' events 
$e^+e^- \rightarrow Z \gamma$ are suited to the study of nonstandard 
physics, particularly if the vector bosons are emitted into 
the central detector region. An effective vertex is constructed 
which contains the most general gauge invariant $e^+e^-Z\gamma$ interaction 
and its phenomenolgocial consequences are examined. 
Low Energy Constraints on the effective vertex are discussed 
as well.

\newpage

{\bf 1. Introduction} 

\vskip2mm

In the last decade a huge number of $Z$'s has been
produced by the LEP1 experiment working on the $Z$--pole and 
the data obtained has been 
used for high precision tests of the Standard Model and to establish stringent 
bounds on physics beyond the Standard Model. Subsequently, the  
LEP2 experiment has started data taking at higher energy with the 
primary aim of determining the mass and selfinteractions of the W--bosons. 
Unfortunately, only a few thousand W--pairs will be available for this 
study, because 
cross sections at LEP2 are generally much smaller than at LEP1, and 
the bounds on new physics will be correspondingly weak \cite{hosm}. 

On the other hand, LEP2 is collecting a relatively large number 
of events with a very hard photon and an on--shell $Z$ in the 
final state which in the Standard Model are produced by 
Bremsstrahlung of the photon from the $e^+e^-$ legs.  
From the experimental point of view the production of $Z$'s 
together with a hard prompt photon is a very clear and pronounced 
signature. Nevertheless,  
this class of events is usually not 
considered very interesting \cite{hosm}, because they seem   
to lead back to LEP1 physics ('radiative return' to on--shell 
$Z$ production). However,  
it is expected \cite{private} that (very roughly) about 
10000 $Z\gamma$ events will be collected until the end of LEP2,   
with an angle $\theta$ of the photon 
larger than 5 degrees with respect to the 
$e^+e^-$ beam direction. 
There will be less of such events 
in the central region $\theta > 30^o$ of the detector, but still 
about 1000 of them will be available to the analysis. 
The primary motivation for the present study is whether at all there 
is a possibility to use these events in a nonstandard physics discussion. 

In our approach we shall use an effective vertex for the 
$e^+e^-Z\gamma$ interaction which contains the Standard Model 
part plus a small admixture of a new contribution. 
The new contribution will be presented in its most general form, i.e. 
as a sum of independent spinor and tensor structures.   
The approach comprises contributions 
from operators of arbitrary high dimensions. 
It can also be considered to effectively 
describe the exchange of new heavy particles 
or some other exotic mechanism for $\gamma Z$ production 
like $e^+e^- \rightarrow Z^\ast \rightarrow Z\gamma$ 
\cite{boudjema}. 

The new contribution is supposed to be small, of the order 
of some small coupling constant $\delta << 1$. Therefore,   
we will be mainly interested in interference terms 
between the Standard Model and the new vertices.
More precisely, if $\sigma_{SM}$ is the Standard Model 
contribution to any differential cross section and 
$\delta \sigma_{NEW}$ is the interference contribution, 
we shall consider the ratio $\delta \sigma_{NEW} / \sigma_{SM}$. 
To a first approximation it is sufficient to use the lowest 
order Standard Model result $\sigma_{LOSM}$ with an on--shell $Z$ 
in the calculation of that ratio. We assume that the Standard Model 
Radiative Corrections are sufficiently small so that only the 
interference amplitude between the Standard Model Contribution and the 
Radiative Corrections are relevant. This interference amplitude can be
recast in terms of form factors we will introduce later on. New Physics 
can appear as form factors which do not arise in the Standard Model 
interference amplitude or as unexpected values for form factors 
which do arise in the Standard Model 
interference amplitude.

\vskip1cm

{\bf 2. Construction of the $e^+e^-Z\gamma$ vertex} 

\vskip2mm

\begin{figure}     
\begin{center}     
\epsfig{file=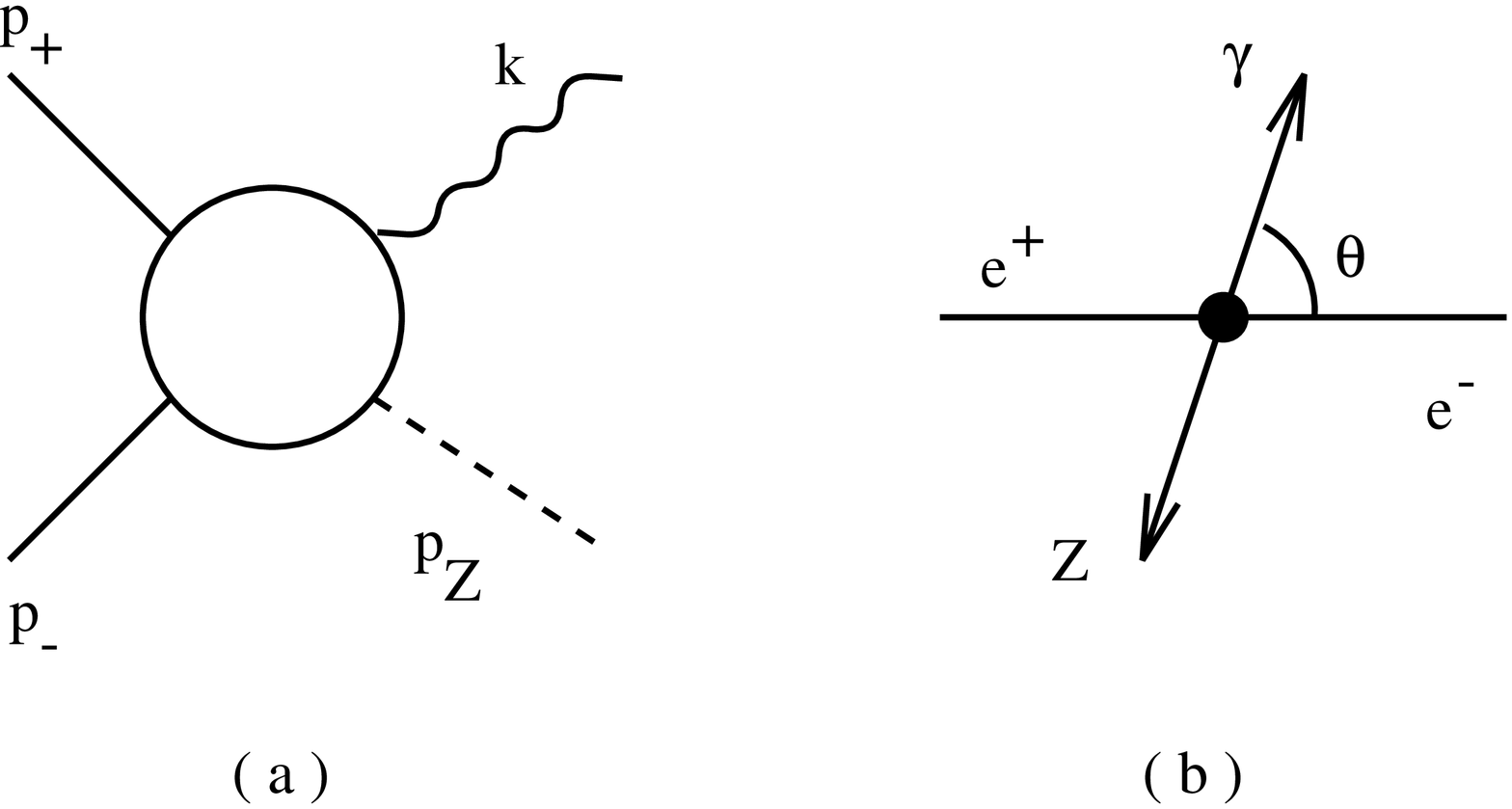,height=9cm,angle=270}
\bigskip           
\caption{               }
\label{fig1}      
\end{center}       
\end{figure}

The process $e^+(p_+) e^-(p_-) \rightarrow \gamma(k) Z(p_Z)$ 
is depicted as a Feynman diagram 
in Fig. \ref{fig1}a and in a kinematic view 
in Fig. \ref{fig1}b. The polarization indices of the 
photon and of the $Z$ are denoted by $\alpha$ and $\mu$, 
respectively. We parametrize the momenta as being 
$p_\pm ={\sqrt s \over 2} (1,0,0,\pm 1)$, 
$k = E_\gamma(1,0,\sin \theta,\cos\theta)$ and 
$p_Z=p_+ + p_- -k$. 
For an on--shell $Z$ ($p_Z^2=m_Z^2$) one obtains a hard monochromatic photon 
of (normalized) energy 
\begin{equation} 
x_\gamma\equiv {E_\gamma\over\sqrt s /2} 
=1-{m_Z^2 \over s}
\label{eq00}
\end{equation}
which is about $x_\gamma=$0.75--0.8 at LEP2 energies 
and goes up to almost 0.97 
at an $e^+e^-$ collider with $\sqrt s =500$ GeV. 
As discussed above, an on--shell $Z$ is a reasonable 
approximation to study nonstandard effects. 
$E_\gamma$ being constant, the process'
real kinematical variable is the production angle $\theta$ of 
the photon. There are essentially 2 regions for photon detection, 
depending on the polar angle,
\begin{itemize}
\item
The region collinear to the beam with large bremstrahlung contributions
is dominated by the Standard Model amplitude which has 
poles at $(k-p_\pm)^2=m_e^2$. 
\item
The central region of the detector, where 
the Standard Model cross section has its minimum 
value, so that one may be sensitive to nonstandard physics. 

\end{itemize}

The amplitude for $Z\gamma$ production has the general form 
$\Gamma_{\mu \alpha}\epsilon_\mu^Z\epsilon_\alpha^\gamma$ where 
$\epsilon_\mu^Z$ is the polarization vector of the $Z$ and 
$\epsilon_\alpha^\gamma$  is the polarization vector of the 
photon. The vertex $\Gamma_{\mu \alpha}$ may be decomposed as 
\begin{equation} 
\Gamma_{\mu \alpha}=\Gamma_{\mu \alpha}^{SM} +\delta\Gamma_{\mu \alpha}
\label{eq1}
\end{equation}
The first term in this expression is the Standard Model contribution 
to the vertex 
\begin{equation} 
\Gamma_{\mu \alpha}^{SM}= - i e^{2} Q_l
\{ \gamma_{\alpha} \frac{k\hspace{-4 pt}/ - p\hspace{-4 pt}/_+ }
                        {(k-p_+)^2-m_e^2}
     \gamma_{\mu} (v_l + a_l\gamma_{5})
  -\gamma_{\mu} (v_l + a_l\gamma_{5}) \frac
              { k\hspace{-4 pt}/ - p\hspace{ -4 pt}/_-}
              {(k-p_-)^2-m_e^2} 
                                     \gamma_{\alpha} \}
\label{eq2}
\end{equation} 
where $a_l={1\over 4 s_W c_W}\approx 0.594$ and 
$v_l=-{1\over 4 s_W c_W}+{s_W\over c_W} \approx -0.05$ 
are the vector and axial
couplings of the electron to the Z. Note that the numerical value 
of $v_l$ is small as compared to $a_l$. This will be important 
later on, when interference terms between  
$\Gamma_{\mu \alpha}^{SM}$ and $\delta\Gamma_{\mu \alpha}$ will 
be discussed and terms of the order $v_l \delta$ will be neglected 
as compared to terms of order $a_l \delta$.  
The second term in Eq. (\ref{eq1}) has the general form 
\begin{eqnarray}
\delta \Gamma_{\mu \alpha} & = &
    ie\{   
      {1\over s} f_1
                     k\hspace{-4 pt}/ \gamma_{\alpha}\gamma_{\mu}
\nonumber \\
    & &  + {1\over s} f_2
                  ( \gamma_{\alpha}k_{\mu} 
                        - k\hspace{-4 pt}/ g_{\alpha\mu}  )
\nonumber \\
   & &  + {1\over s^2} f_3 \gamma_{\mu}
                 (p_{+,\alpha} kp_- - p_{-,\alpha} kp_+ )
\nonumber \\
   & &   + {1\over s^2} f_4 p_{-,\mu}
                 (\gamma_{\alpha} kp_+ - p_{+,\alpha}k\hspace{-4 pt}/  )
     + {1\over s^2} f_5 p_{+,\mu}
                 (\gamma_{\alpha} kp_+ - p_{+,\alpha}k\hspace{-4 pt}/  )
\nonumber \\
   & &   + {1\over s^2} f_6 p_{-,\mu}
                 (\gamma_{\alpha} kp_- - p_{-,\alpha}k\hspace{-4 pt}/  )
     + {1\over s^2} f_7 p_{+,\mu}
                 (\gamma_{\alpha} kp_- - p_{-,\alpha}k\hspace{-4 pt}/  )
                             \}
\label{eq3}
\end{eqnarray}
where $f_i << 1$ are dimensionless coupling constants whose strength 
cannot be predicted within our approach. One factor of $e$ has been 
introduced in the definition of the vertex (\ref{eq3}) for convenience. 
In writing down this formula several requirements are taken care of: 
\begin{itemize}
\item The requirement of electromagnetic gauge invariance is fulfilled 
      by forming in Eq. (\ref{eq3}) suitable combinations such that 
      $k^\alpha \delta \Gamma_{\mu \alpha}  =0$. 
      under infinitesimal gauge transformations. 
\item Eq. (\ref{eq3}) is not explicitly $SU(2)_L$ invariant. 
      The point is that in our approach the new interactions 
      are not necessarily tied to very high energies but could 
      in principle be related to energies below 1 TeV. 
      Therefore we have refrained from make Eq. (\ref{eq3}) 
      explicitly $SU(2)_L$ invariant. However, if desired, 
      one can enforce global $SU(2)_L$ by adding a similar 
      correction to the $e\nu \gamma W$ vertex. This will enforce 
      additional low--energy constraints on the new vertex to 
      be discussed in section 4. 

      It should be noted that our vertex is trivially invariant 
      under those local $SU(2)_L$ gauge variations which 
      transform the $Z$ into itself, because these are given 
      by $\delta \epsilon^\mu_Z \sim p_Z^\mu$ and thus one needs 
      to have $p_Z^\mu \delta \Gamma_{\mu \alpha}  =0$. 
      Although this condition is not explicitly fulfilled by 
      Eq. (\ref{eq3}), the situation is automatically cured, because 
      one can replace 
      all 4--vectors $q={p_-,p_+,k}$ by 
      $q_\mu-{q p_Z \over m_Z^2}p_{Z,\mu}$ and $g_{\alpha\mu}$   
      by $g_{\alpha\mu}-{p_{Z,\mu}p_{Z,\alpha} \over m_Z^2}$, and  
      this replacement does not change the cross section, 
      because terms $\sim p_{Z,\mu}$ give $O(m_e)$ when sandwiched 
      between the lepton spinors.
\item
      The 'coupling constants' $f_i$ are really form factors 
      $f_i=f_i(x_\gamma, \cos\theta)$. Powers of $s$ have been 
      introduced in Eq. (\ref{eq3}) in such a way as to make 
      the $f_i$ dimensionless. 
      The new contribution would have the same overall energy 
      dependence as the Standard Model contribution, if one would 
      assume the $f_i$ to first approximation to be constant in 
      energy $x_\gamma=1-m_Z^2/ s$. However, this would be a rather 
      unpleasant feature, because it would induce effects already 
      at the lowest energies (see the discussion of low energy 
      constraints in section 4). 
      Therefore, it is a good idea to assume that the form factors 
      behave like some negative power of $1-x_\gamma$ in order to cut away 
      the low energy constraints. Physically, such a behavior 
      arises, for example, if the new physics is induced at some 
      scale $\Lambda \sim 1$ TeV 
      which forces the form factors to behave like a power of 
      $s/\Lambda^2$. Furthermore, such behaviour is also desirable for
      the purpose of satisfying constraints from unitarity at high 
      energies. 
\item
      The coupling constants $f_i$ are of the general 
      form $f_i=v_i+a_i \gamma_5$ and are really form factors 
      $f_i=f_i(x_\gamma, \cos\theta)$. 
\item
      Terms which give contributions of order $m_e /\sqrt s$ 
      when the interference with the Standard Model is formed, 
      have not been included in Eq. (\ref{eq3}). Such terms 
      consist of a product of an even number of $\gamma$ matrices. 
 \item  
    The interactions in Eq. (\ref{eq3}) conserve CP. 
    Using complex form factors 
    one could also undertake a search for CP violating interactions -- 
    in analogy to what has been done in ref. \cite{lampeabraham} 
    concerning the $bb\gamma Z$ vertex. 
\end{itemize}

\vskip1cm

{\bf 3. Quantitative Phenomenological Consequences} 

\vskip2mm

\begin{figure}
\begin{center}
\epsfig{file=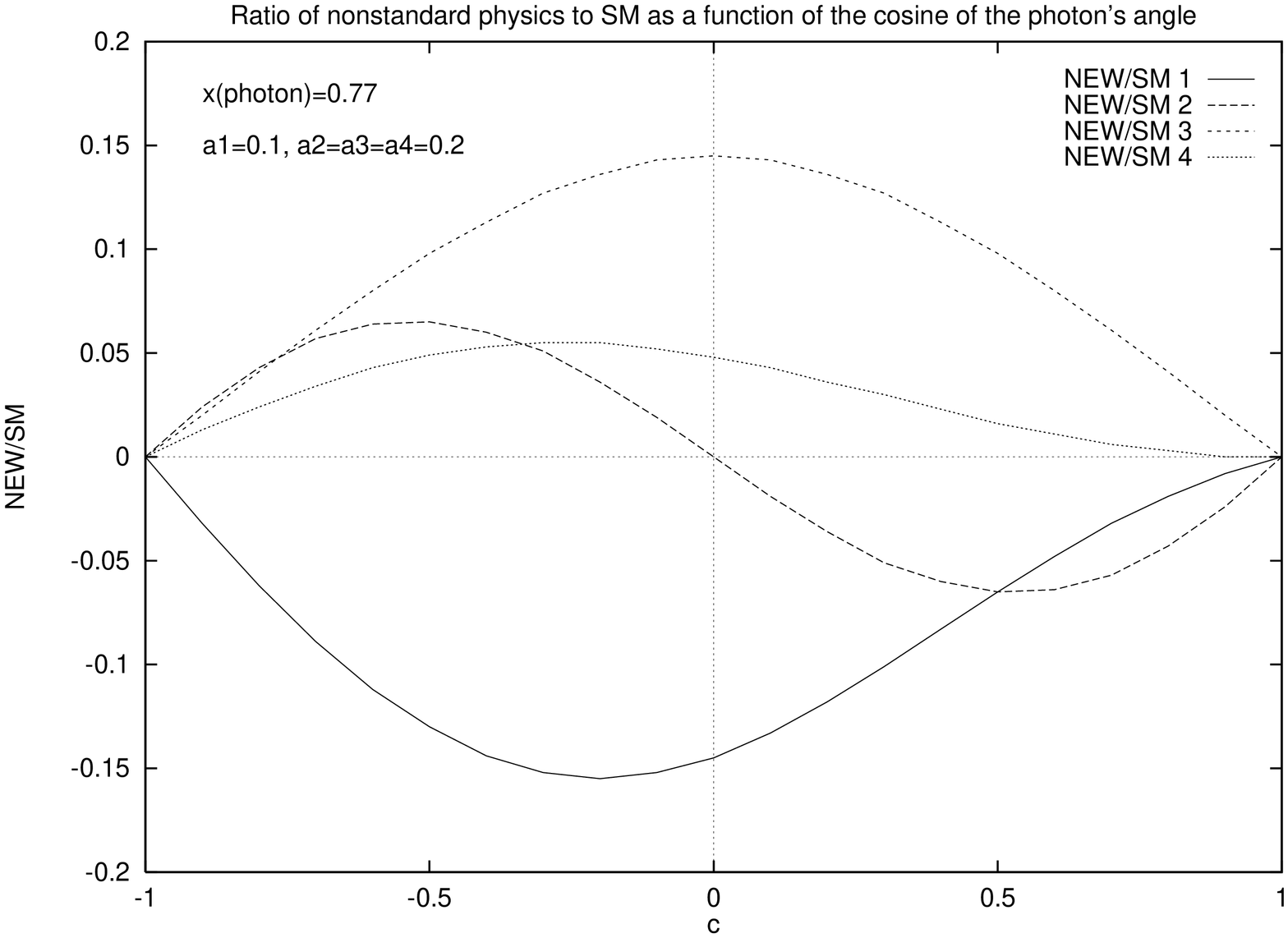,height=10cm,angle=270}
\bigskip
\caption{               }
\label{figat}
\end{center}
\end{figure}

With the vertex (\ref{eq3}) at hand one can calculate  
the cross section $d\sigma / d\cos\theta$ where $\theta$ 
is the production angle of the photon. 
$d\sigma / d\cos\theta$ 
consists of a Standard Model term $d\sigma_{SM} / d\cos\theta$ 
which is proportional to $v_l^2+a_l^2$ and 
an interference term $d\sigma_{NEW} / d\cos\theta$. 
      These interference contributions between 
      $\Gamma_{\mu \alpha}^{SM}$ and $\delta\Gamma_{\mu \alpha}$  
       are $\sim v_l v_i -a_l a_i$.  
For convenience the numerical analysis will be done 
only for the form factors $f_{1,2,3,4}$ but not for $f_{5,6,7}$.
       Due to the fact that  
      the Standard Model coupling $v_l$ almost vanishes it turns 
      out that the $v_i$ practically do not contribute to the interference 
      term and that the ratio of $d\sigma_{NEW} / d\cos\theta$ 
  and $d\sigma_{SM} / d\cos\theta$ is proportional to $a_i/a_l$. 
Explicitly one has   
\begin{eqnarray}
{d\sigma_{NEW} / d\cos\theta \over d\sigma_{SM} / d\cos\theta} & =&
{1 \over v_l^2+a_l^2}
{1 \over 4(x_\gamma^2+2-2x_\gamma)/\sin^2\theta -2}
\nonumber \\
& & \{  (v_l v_1 -a_l a_1)(2-x_\gamma (1+\cos\theta) )
\nonumber \\
& &  +(v_l v_2 -a_l a_2) x_\gamma \cos\theta
\nonumber \\
& & +(v_l v_3 -a_l a_3) (2x_\gamma -1)
\nonumber \\
& & +(v_l v_4 -a_l a_4) x_\gamma (\cos\theta-1)    + \cdots       \}
\label{eq4}
\end{eqnarray}
where $x_\gamma=1-m_Z^2/s$ as before and 
the dots stand for similar terms stemming from the other 
formfactors $i > 4$. Note that  $d\sigma_{SM} / d\cos\theta$
becomes very large in the collinear regions $\cos\theta\approx\pm 1$ 
and stays roughly constant and small in the central region 
$\cos\theta\approx 0$. Fig. \ref{figat} shows the ratio 
(\ref{eq4}) as a function of $\cos\theta$ for form factors 
with numerical values $a_1=0.1$, $a_2=0.2$, $a_3=0.2$ and $a_4=0.2$ 
at $x_\gamma=0.77$ (corresponding to $\sqrt s =190$ GeV).    
It is nicely seen that the different form factors contribute 
quite differently to the $\theta$ distribution. 
The knots at $\cos\theta =\pm 1$ are due to the fact that 
the new contributions remain regular at this point whereas 
the Standard Model cross section is large. Note that the result 
is linear in the $a_i$ and that the 
order of magnitude of the result is roughly the same as 
the magnitude of the $a_i$.  

\begin{figure}
\begin{center}
\epsfig{file=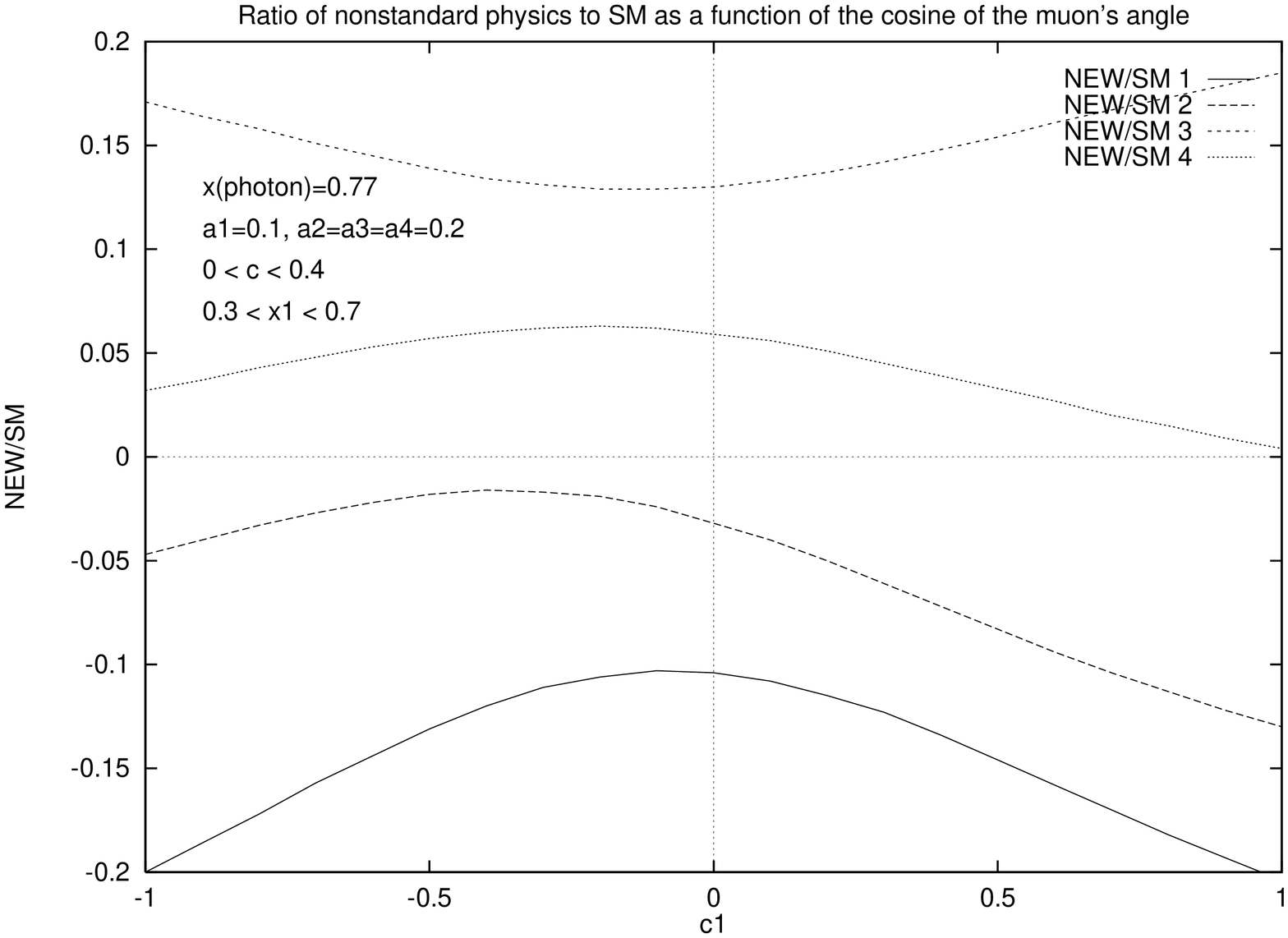,height=10cm,angle=270}
\bigskip
\caption{               }
\label{figac1} 
\end{center}
\end{figure}

One can learn more about the nature of the process 
$e^+ e^- \rightarrow \gamma Z$ if one 
analyzes correlations with the decay products of the $Z$, e.g. 
muons. One should not, however, give up the condition 
of an on--shell $Z$. This is because otherwise a lot of 
other Standard Model processes like 
$e^+ e^- \rightarrow Z^\ast \rightarrow \mu^+ \mu^- \gamma $ 
would contribute and complicate the analysis. Experimentally,  
it is straightforward to isolate a sample with on--shell $Z$'s 
by cutting in the $x_\gamma$ distribution, because the 
$x_\gamma$ distribution reflects the $Z$ resonance.
If one restricts to a bin in the neighbourhood of the 
pole at $x_\gamma=1-m_Z^2/s$, there are only small irrelevant off-shell 
corrections as discussed in detail in the introduction. 

A second important issue is that the 
$Z\rightarrow\mu^+\mu^-$ vertex is assumed to be that of the Standard Model.
This assumption is justified as any deviation is constrained from Lep-1 
data to be small.
Therefore, from $Z$ decay one has factors of the form $v_m^2+a_m^2$ 
or $v_ma_m$
in all terms of the complete matrixelements, where $v_m$ and $a_m$ are 
vector and axial--vector coupling of the $Z$ to its decay product  
(lepton or quark).

\begin{figure}
\begin{center}
\epsfig{file=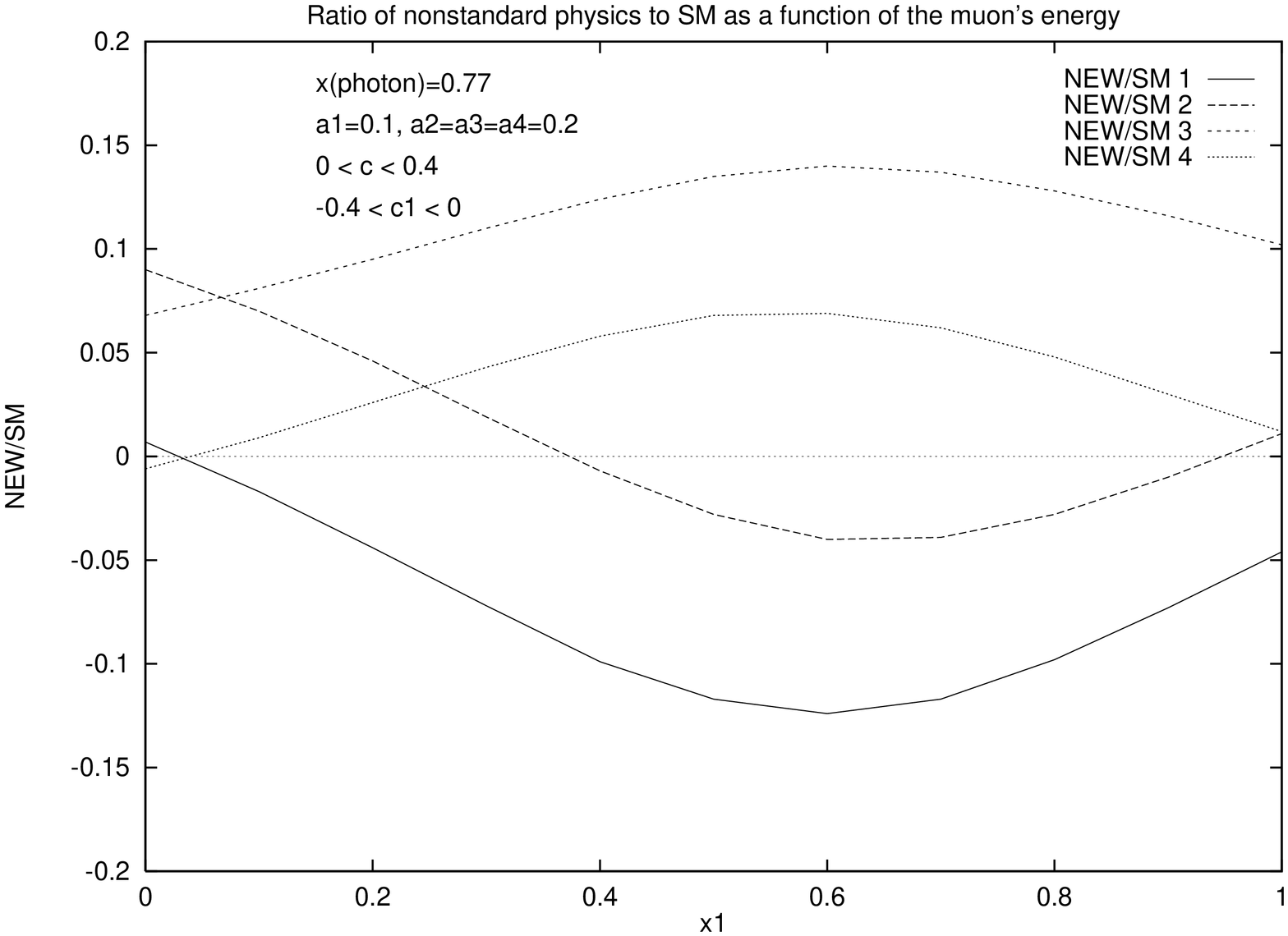,height=10cm,angle=270}
\bigskip
\caption{               }
\label{figax1} 
\end{center}
\end{figure}

Let us now discuss in detail some distributions of $Z$ decay products, 
assuming first that the $Z$ decays leptonically.  
As without Z--decay, all ratios  
$d\sigma_{NEW} /d\sigma_{SM}$ behave like $a_i/a_l$ 
to a good approximation.  
The reason for that is mainly due to $v_m\approx 0$ for leptons. 
The Standard Model terms 
either go with $(v_l^2+a_l^2)(v_m^2+a_m^2)\approx a_l^4$ or with 
$(v_la_l)(v_ma_m)\approx 0$ where one factor in the squares is due 
to production of the Z and the other factor is due to the decay. 
The interference terms either go with $(v_iv_l-a_ia_l)(v_m^2+a_m^2)
\approx a_ia_l^3$ or $(v_ia_l-a_iv_l)v_ma_m\approx 0$, so that 
$d\sigma_{NEW} /d\sigma_{SM} \sim a_i/a_l$ as claimed. 

In order to be as sensitive as possible to new physics contributions,  
a sample of events with photons in the central region of the 
detector say $0< \cos\theta < 0.4$ should be chosen. 
We have taken an asymmetric bin (i.e. no event with $\cos\theta < 0$) 
because some form factors yield contributions asymmetric in 
$\cos\theta$, as seen in Fig. \ref{figat}. Next we assume that the  
Z decay to $f\bar f$ with momentum $p_f={\sqrt{s} \over 2} x_1 
(1,\sin\phi_1\sin\theta_1,\cos\phi_1\sin\theta_1,\cos\theta_1)$ 
in the lab system. One can then 
study the energy ($x_1$) and angle 
($c_1:=\cos\theta_1$) dependence.
Ratios $d\sigma_{NEW} /d\sigma_{SM}$ have been plotted as a function 
of these variables in Figs. \ref{figac1} and \ref{figax1}. 
The same values of couplings $a_1=0.1$, $a_2=0.2$, $a_3=0.2$ and $a_4=0.2$ 
and the same energy $x_\gamma=0.77$ as in Fig. \ref{figat} has been 
chosen. Furthermore, we have averaged over the bin $0< \cos\theta < 0.4$ 
and in Fig. \ref{figac1} in addition over $0.3<x_1<0.7$ and 
in Fig. \ref{figax1} over $-0.4 < c_1 <0$.  

\vskip1cm

{\bf 4. Summary and Discussion} 

\vskip2mm

In this letter possible new physics contributions to 
the LEP2 process $e^+e^- \rightarrow Z \gamma$ have been 
analyzed by an effective vertex ansatz. Several nontrivial 
features of the new interactions have been derived. 
One may ask the question why we  
did not use the fashionable effective Lagrangian 
approach, in which new interactions are expanded in powers 
of higher dimensional operators, in particular dimension 6, 
and added to the Standard Model Lagrangian. Such operators 
are preferably chosen to respect the Standard Model $SU(2)_L\times U(1)_Y$ 
gauge sysmmetries. The complete set of these operators 
inducing the process $e^+e^- \rightarrow Z \gamma$ is given by 
\begin{itemize}
\item $\bar l D_\mu e D^\mu \phi    \qquad \qquad \qquad \qquad
             D_\mu\bar l  e D^\mu \phi  $ 
\item 
$\bar l \sigma_{\mu\nu}\tau^a e \phi  W_{\mu\nu}^a            $ 
\item 
$i\bar l \tau^a \gamma_\mu  D_\nu l   W_{\mu\nu}^a  \qquad \qquad \qquad 
  i\bar l  \gamma_\mu  D_\nu l   B_{\mu\nu}   $ 
\item $i\bar e \gamma_\mu  D_\nu  B_{\mu\nu} e  $
\end{itemize}
where $e$ and $l$ denote the righthanded electron and left handed 
lepton doublet respectively. $\phi$ is the Higgs doublet with 
vacuum expectation value $<\phi >=(0,v)$ and $W_{\mu\nu}^a$ and 
$B_{\mu\nu}$ are the $SU(2)$ and $U(1)$ field strengths. 
Note that in all the cases the $e^+e^- Z \gamma$ interaction 
is induced indirectly as a higher order effect either by the 
gauge field in a covariant derivative $D_\mu$ or by the nonlinear term 
in the nonabelian field strength. This implies that all these 
dimension 6 interactions induce $e^+e^- \gamma$ or $e^+e^- Z$ 
couplings much stronger than the $e^+e^- \gamma Z$ coupling.  
For all of them therefore exist much stronger constraints 
from LEP1 than from LEP2. 
\footnote{Actually, the first 3 operators above flip the 
helicity and therefore contribute only 
$O(m_e)$ to all cross sections.}  
This is the reason why we had to do without the effective 
Lagrangian approach to study new physics effects in 
$e^+e^- \rightarrow Z \gamma$.
In our approach we avoided the restriction to dimension 6, 
because the effective vertex in principle collects contributions 
from operators of arbitrary high dimension. For example, 
the first form factor $f_1$ gets a contribution from a 
dimension 8 operator of the form 
$\bar l \gamma^\mu l \epsilon_{\mu\nu\sigma\tau} 
   D^\nu B^{\sigma\lambda} B_\lambda^\tau$ 
first studied in ref. \cite{parke}.  

Now we come to the question of constraints from lower energies. 
As has already been discussed, bounds from really low energies 
can be avoided by assuming a suitable energy dependence of 
the form factors of the form $f_i \sim s$. In that 
case the only potentially important experiment to consider 
is LEP1 with an energy only a factor 2 below LEP2. 
At LEP1, events of the form 
$Z \rightarrow f\bar f\gamma$ can be described by our vertex.  
With an energy dependence $f_i \sim s$ the expected 
ratios $\delta \sigma_{NEW} / \sigma_{SM}$ are roughly a factor 
of 4 smaller than those considered in section 3.   
This means for a comparable study LEP1 would 
need about $4\times 10000$ events of 
the type $Z \rightarrow f\bar f\gamma$. 
LEP1 has obtained a lot of photon events and analyzed them 
in various studies. However, the photons of interest here 
are hard and noncollinear, and there are only of the order of 
thousand such events in each detector at LEP1. 
They have been used by the L3 group  
\cite{lep1} to derive constraints on $ZZ\gamma$ and 
$Z\gamma\gamma$ couplings and could in principle be used 
to obtain limits on our form factors as well.  

If one considers our vertex (\ref{eq3}) as part of a $SU(2)_L$ 
symmetric system, one has in principle constraints from 
lepton--neutrino scattering processes $l\nu\rightarrow W\gamma$. 
However, apart from being done at rather low energies, 
lepton--neutrino scattering has much too small cross sections 
to compete with $Z\gamma$ production at LEP2. 

Another constraint could come from Tevatron data 
$\bar p p\rightarrow Z\gamma X$ if one assumes in the spirit 
of quark--lepton universality that light quarks have similar 
anomalous couplings to $Z\gamma$ as electrons.
This process is well studied in conjuction with a 
parallel analysis of $\bar p p\rightarrow W\gamma X$ which
constrains the 
the triple $WW\gamma$ gauge coupling via 
$\bar u d \rightarrow W^\ast\rightarrow Z\gamma$. 
The authors of ref \cite{tevatron} have given a 
exhaustive review about the results from D0 and CDF. 
It turns out that there are less than 100 candidate 
events of the type $\bar q q\rightarrow Z\gamma$ if one 
assumes that the $Z$ decays leptonically. This latter 
restriction is important because the background in the 
hadronic channel is too large. The number of events is  
too low to compete with the statistics of the 'radiative 
return' events at LEP2. The ability to test 
nonstandard physics is further reduced, because the photon transverse 
momenta are not really large, typically $E_T^\gamma > 10$ GeV.  
Furthermore, the form of the $E_T^\gamma$ distribution is not as sensitive 
to the structure of new physics as the $\cos\theta$ distribution 
which was considered for LEP2. 

Finally we want to stress that the main aim of this letter is 
to show that one can construct new physics possibilities for the LEP2 process 
$e^+e^- \rightarrow Z \gamma$ in spite of various obstacles 
like gauge invariance, dominance of low dimensional operators, 
LEP1 constraints etc. Our main conclusion is therefore that it is 
worthwhile to analyze these events as precise as possible
and that one should not look at them just as boring 
background from the Standard Model. 

\vskip1cm

{\bf Acknowledgements:} 
We are indebted to G\"unter Duckeck and Ron Settles, who have  
informed us about important experimental aspects.  
Arnd Leike has triggered our analysis of low energy  
constraints. 

\vskip2mm


\end{document}